\documentstyle[10pt,emulateapj]{article}

\received{December 15 2000}
\accepted{January 22 2001}

\begin{document}

\title{Constraints on the warm dark matter model 
from gravitational lensing} 

\author{Yan-Jie Xue and Xiang-Ping Wu}

\affil{Beijing Astronomical Observatory, Chinese Academy
                 of Sciences, Beijing 100012; 
       and National Astronomical Observatories, Chinese Academy
                 of Sciences, Beijing 100012; China}

\begin{abstract}
Formation of sub-galactic halos is suppressed in warm dark matter (WDM) 
model due to thermal motion of WDM particles. 
This may provide a natural resolution to some puzzles in 
standard cold dark matter (CDM) theory such as
the cusped density profiles of virialized dark halos and the overabundance 
of low mass satellites. One of the observational tests of the WDM model
is to measure the gravitationally lensed images of distant quasars
below sub-arcsecond scales. In this {\sl Letter}, we report a comparison
of the lensing probabilities of multiple images between CDM and WDM models 
using a singular isothermal sphere model for the mass density profiles of 
dark halos and the Press-Schechter mass function for their distribution 
and cosmic evolution. It is shown that the differential probability of 
multiple images with small angular separations down to $\sim10$ 
milliarcseconds should allow one to set useful constraints on the WDM 
particle mass. We discuss briefly the feasibility and uncertainties of
this method in future radio surveys (e.g. VLBI) for gravitational lensing. 
\end{abstract}


\section{Introduction}

While cosmological models with a mixture of roughly $35\%$ cold dark matter
(CDM) and $65\%$ vacuum energy have proved remarkably successful at explaining
the origin and evolution of cosmic structures on large-scales ($>1$ Mpc),
there are a number of independent observations on small scales
that seem to be in conflict with the predictions of standard
CDM models such as the cusped density profiles of virialized dark halos 
and the overabundance of satellite structures. 
Among many other mechanisms suggested for resolving these discrepancies,
the simplest approach is to smear out small scale power by free-steaming 
due to thermal motion of particles. Warm dark matter (WDM) is thus invoked 
and has recently received a lot of attention in literature   
(e.g. White \& Croft 2000; Col\'in, Avila-Reese \& Valenzuela 2000;
Sommer-Larsen \& Dolgov 2000; 
Bode, Ostriker \& Turok 2000b; references therein).

In the conventional scenario of structure formation by WDM, 
perturbations on small mass scales $M_{\rm f}\approx4\times 10^{11}M_{\odot}
h^2\Omega_{\rm X}(R_{\rm f}/0.1\;{\rm Mpc})^3$ are damped, relative to
CDM models, where $\Omega_{\rm X}$ is the WDM density parameter
and $R_{\rm f}$ denotes the characteristic free-streaming length.
As a result, it suppresses the formation of small halos with 
$M<M_{\rm f}$ at high redshift and meanwhile reduces the overall number of 
low mass halos in the universe, depending sensitively on the WDM 
particle mass $m_{\rm X}$. Observational tests of this scenario
and hence robust constraints on $m_{\rm X}$ are restricted to 
the studies of less massive systems with $M<M_{\rm f}$. 
The existing limits in literature are based on the  Gunn-Peterson effect  
and the Ly-$\alpha$ forest, which
set $m_{\rm X}>750$ eV (Narayanan et al. 2000).
In this {\sl Letter}, we would like to examine whether the gravitational
lensing of high-redshift quasars by foreground dark halos  
can be used to distinguish CDM and WDM models or place any meaningful
constraints on $m_{\rm X}$. This arises because 
the probability of gravitational lensing would be lower in
WDM model, as a result of the suppression 
of formation of small mass halos on galactic and sub-galactic scales, 
than in CDM model. 
For a massive galaxy like the Milky Way as a lens at cosmological distance, 
the Einstein radius of a background quasar is typically 1 arcsecond.
So, it is expected that the difference in lensing properties
between WDM and CDM models would occur below sub-arcsecond scales. 
We would like to demonstrate quantitatively to what angular separations the 
two models become to be distinguishable and how sensitively their 
lensing probabilities depend on the WDM particle mass.

\section{Lensing probability}

For simplicity, we use a singular isothermal sphere to model the mass 
distribution inside a virialized halo so that the total mass $M$ within
radius $r$ is simply $M=2\sigma_{\rm v}^2r/G$. We truncate the mass profile at 
the virial radius $r_{\rm vir}$ defined by 
$M_{\rm vir}=4\pi\Delta_{\rm c}\rho_{\rm c} r_{\rm vir}^3/3$, where
$\rho_{\rm c}=3H^2(z)/8\pi G$ is the critical mass density of the universe,
$H(z)=100\;h\;E(z)$ km s$^{-1}$ Mpc$^{-1}$ is the Hubble constant,
$E(z)=\Omega_{\rm M}(1+z)^3+\Omega_{\rm K}(1+z)^2+\Omega_{\Lambda}$,
$\Omega_{\rm K}$ denotes the curvature term, 
and $\Delta_{\rm c}$ represents the overdensity of dark matter with respect to
the critical density $\rho_{\rm c}$ 
at the collapse redshift $z$. For a dark halo at $z$ as a lens,
the multiple images of a distant quasar at $z_{\rm s}$
with an angular separation
of $\Delta\theta=2\theta_{\rm E}$ will occur if 
it happens to lie in the Einstein radius of the dark halo: 
$\theta_{\rm E}=4\pi(\sigma_{\rm v}/c)^2(D_{\rm ds}/{D_{\rm s}})$, in which
$D_{\rm ds}$ and $D_{\rm s}$ are the angular diameter distances from
the lens and observer to the source, respectively. The lensing cross section 
for the multiple images is simply $\pi \theta_{\rm E}^2$. 
The probability that a quasar at $z_{\rm s}$ will have 
multiple images with an angular separation greater than $\Delta\theta$ 
due to all the foreground dark halos is thus 
\begin{equation}
P(\Delta\theta)=\int_{0}^{z_{\rm s}}\left(\frac{dV}{dz}\right) 
		dz \int_{M_{\rm min}}^{\infty}
	  \left(\frac{\theta_{\rm E}^2}{4}\right)
          \left(\frac{dN}{dMdV}\right)    dM,
\end{equation}
where $dV$ is the comoving volume element, and $M_{\rm min}$ is related to
$\Delta\theta$ through
\begin{eqnarray}
M_{\rm min}=& 1.50\times 10^{12}M_{\odot}\;h^{-1} \nonumber\\
    &     \left(\frac{1}{E(z)}\right)
        \left(\frac{\Delta \theta}{1^{\prime\prime}}\right)^{3/2}
        \left(\frac{D_{\rm s}}{D_{\rm ds}}\right)^{3/2}
        \left(\frac{\Delta_{\rm c}}{200}\right)^{-1/2}.  
\end{eqnarray}

We use the Press-Schechter (PS) mass function for the abundance and
evolution of virialized dark halos:
\begin{equation}
\frac{dN}{dMdV}=-\sqrt{\frac{2}{\pi}} \frac{\bar{\rho}}{M}
    \frac{\delta_c(z)}{\sigma^2}   \frac{d\sigma}{dM}
    \exp\left(-\frac{\delta_c(z)^2}{2\sigma^2}\right),
\end{equation}  
where $\bar{\rho}$ is the present-day average mass density of the
universe,  $\delta_{\rm c}(z)$ is the linear over-density of perturbations 
that collapsed and virialized at redshift $z$, and  $\sigma$ is 
the variance of the mass density
fluctuation in sphere of mass $M=(4\pi/3)\overline{\rho}R^3$. 
It is convenient to express $\delta_{\rm c}(z)$ as 
$\delta_{\rm c}(z)=\delta_0(1+z)[g(0)/g(z)]$, where $g(z)$ is the linear growth
factor, for which we take the approximate formula by Carroll, 
Press \& Turner (1992), and the quantity $\delta_0$ has a weak dependence
on $\Omega_{\rm M}$: $\delta_0=1.6866[1+0.01256\log \Omega_{\rm M}(z)]$
for a flat universe with nonzero cosmological constant (Mathiesen \&
Evrard 1998). Once a power spectrum, $P(k)$, is specified,
the mass variance becomes
\begin{equation}
\sigma^2(M)=\frac{1}{2\pi^2}\int_0^{\infty} k^2 P(k) W^2(kR) dk, 
\end{equation}  
in which $W(x)=3(\sin x-x\cos x)/x^3$  
is the Fourier representation of the window function.  
We adopt the following parameterization of WDM power spectrum 
\begin{equation}
P(k)=A\;k^n\;T_{\rm X}^2(k)\;T_{\rm CDM}^2(k),
\end{equation}  
where $n$ is the primordial power spectrum and is assumed to
be the Harrison-Zeldovich case $n=1$, 
$T_{\rm CDM}(k)$ is the transfer function of adiabatic CDM model
for which we use the fit given by Bardeen et al. (1986), and 
$T_{\rm X}(k)$ acts as the CDM to WDM `` transfer function'' for which
we adopt the form provided by Bode et al. (2000b): 
$T_{\rm X}(k)=[1+(\alpha k)^{2\nu}]^{-5/\nu}$, where $\nu=1.2$ and
$\alpha$ is related to the WDM particle mass $m_{\rm X}$ through
$\alpha=0.048 (\Omega_{\rm X}/0.4)^{0.15} 
(h/0.65)^{1.3} ({\rm keV}/m_{\rm X})^{1.15}$. 
The amplitude $A$ is determined by equation (4) using the rms mass 
fluctuation on an 8 $h^{-1}$ Mpc scale, $\sigma_8$.

We work with a flat cosmological model where 
$\Omega_{\rm M}=\Omega_{\rm X}+\Omega_{\rm b}=0.3$, 
$\Omega_{\rm b}=0.03$, $\Omega_{\Lambda}=0.7$, and $h=0.68$, 
which fixes the shape parameter 
$\Gamma=\Omega_{\rm M}h\exp[-\Omega_{\rm b}(1+\sqrt{2h}/\Omega_{\rm M})]
=0.176$.
The normalization parameter is taken from the calibration of
cluster abundance, $\sigma_8=0.85$. 
We compute the lensing probabilities for 
a quasar at redshift $z_{\rm s}=1$, $2$ and $3$, respectively,
and for two choices of the WDM particle mass $m_{\rm X}=350$ eV and 1 keV.
As a comparison, we also give the corresponding lensing probabilities
for a CDM model with $\Omega_{\rm CDM}=0.27$.  In order to highlight 
their differences, we display in Figure 1 the normalized differential 
probabilities 
$[1/P(\Delta\theta)][dP(\Delta\theta)/d\Delta\theta]$ instead of the
total probabilities $P(\Delta\theta)$. For the latter, the typical
value for a quasar at  $z_{\rm s}=2$ with $\Delta\theta>1^{\prime\prime}$
is nearly the same  for all the three 
models (two WDM models and the standard CDM model):
$P(>1^{\prime\prime})\approx1\times10^{-3}$. This result is   
consistent with the pioneering statistical study of gravitational lensing 
by Turner, Ostriker \& Gott (1984). 
Noticeable departure of the lensing probability
in WDM models from that in CDM model occurs only at small angular 
separations $\Delta\theta<1^{\prime\prime}$ and 
$\Delta\theta<0.^{\prime\prime}1$ for  $m_{\rm X}>350$ eV and
$m_{\rm X}>1$ keV, respectively. It is likely that one needs to reach an
even smaller angular separation of  $\sim10$ milliarcseconds
in order to make an effective distinction between the WDM and CDM models
for a realistic WDM particle mass of $m_{\rm X}\sim1$ keV. 
We have applied the lensing probability $P(\Delta\theta)$ to 
the quasar luminosity function determined from the 2dF QSO survey
over absolute magnitudes $-26<M_{\rm B}<-23$ and  
redshifts $0.35<z_{\rm s}<2.3$ (Boyle  et al. 2000).
Here we would rather take a less vigorous approach to the problem by 
neglecting the magnification effect of the double images. 
It is shown that at a limiting magnitude of $m_{\rm B}=20$,  
the fractions of small angular separation events
with $0.^{\prime\prime}001<\Delta\theta<0.^{\prime\prime}01$ 
in the total number of doubly-imaged quasars 
are approximately $1/100$, $4/1000$ and $1/1000$ for the CDM model and 
the WDM models with $m_{\rm X}=1$ keV and $m_{\rm X}=350$ eV, respectively.
Indeed, observational tests of these fractions on
milliarcsecond scales turn out to be difficult and challenging.

\section{Discussion and conclusions}

In the CDM scenario of structure formation, many small galaxies
would form, relative to what are predicted by the WDM
model. Consequently, the standard CDM model would predict  
more multiple, small angular separation images of the gravitationally lensed 
distant quasars than WDM model does. This unique property can be used 
as an effective tool to distinguish these two prevailing models. 
It has been shown that the significant difference in the 
distributions of image angular separations
between WDM and CDM models  occurs at small angular separations
$\Delta\theta\sim0.01^{\prime\prime}$ for the WDM particle mass 
$m_{\rm X}\sim1$ keV. Although there are no data available
at present to test our prediction, these angle ranges should be 
accessible by current radio observations such as VLBI. Recall that systematic 
radio surveys for strong  gravitational lensing have proved to be
very successful at finding small separation images 
in the range $0.^{\prime\prime}3$-$0.^{\prime\prime}6$ 
(e.g. Augusto, Wilkinson \& Browne 1998; Augusto \& Wilkinson 2001; and
references therein). It is hoped that future multiple image surveys down to
$\sim10$ milliarcseconds would allow one to place useful constraints on
the WDM particle mass.

We now briefly discuss several major uncertainties in the present 
predictions:
Firstly, we have used a simple singular isothermal sphere model 
for the mass distributions of dark halos. 
An alternative is the universal density profile 
suggested by high-resolution $N$-body simulations 
(Navarro, Frenk \& White 1995; NFW), which should be
valid for the virialized dark halos as small as $10^{9}M_{\odot}$.
Li \& Ostriker (2000) have recently compared the lensing probabilities
$P(\Delta\theta)$ produced by a singular isothermal sphere model 
and the NFW profile using roughly the same approach as ours. 
It turns out that the NFW profile results in a significantly small
value of $P(\Delta\theta)\sim 10^{-6}$,  
insensitive to the image separation $\Delta\theta$.
We have recalculated the lensing probabilities under CDM and WDM models 
following the approximate treatment of Li \& Ostriker (2000) 
for the image separation by NFW profile,
$\Delta\theta\approx 2\theta_0$ where $\theta_0$ is the solution 
to the lensing equation with the zero alignment parameter.  
Unlike Li \& Ostriker (2000) who took a 
constant halo concentration $c$ for massive halos from numerical
simulations,  we fix the 
characteristic density $\delta_{\rm c}$ and halo concentration $c$ 
for each halo $M$ using the $\delta_{\rm c}$-$M$ relation 
via the collapse redshift $z_{\rm coll}$ proposed
by Navarro, Frenk \& White (1997). Nevertheless, we have reached 
a conclusion essentially similar to Li \& Ostriker (2000). 
For example,  the lensing probabilities for a distant quasar at $z_{\rm s}=3$
with $\Delta\theta\geq 0.001^{\prime\prime}$ -- $1^{\prime\prime}$ 
are $P(\Delta\theta)\approx 
3.7\times10^{-6}$, $3.6\times10^{-6}$ and $2.9\times10^{-6}$ for
the CDM model and the WDM models 
with $m_{\rm X}=1$ keV and $m_{\rm X}=350$ eV, 
respectively. The lack of small separation events below arcsecond scales 
will invalidate our original intention that the study of 
doubly-imaged quasars may help to distinguish between CDM and WDM models. 
However, we note that the lensing probability produced by NFW profile 
is unreasonably small ($\sim10^{-6}$). This implies that the NFW profile
may be inappropriate to the description of the mass distributions in
the inner regions of dark halos where the multiple images would occur.
Indeed, the central mass density of an NFW profile is too shallow 
to act as an effective lens. 
The absence of detectable odd images in the known lens systems 
also lends support to the existence of 
a steeper inner density profile in galaxies,
$\rho\propto r^{-2}$ (Rusin \& Ma 2001). In a word, 
the singular isothermal sphere model may be 
a reasonable approximation for the inner mass profiles of galactic and
sub-galactic halos.

Secondly, the PS mass function may become inaccurate
for small halos. High resolution $N$-Body simulations have shown that
the PS approximation predicts too many small halos at
low redshifts (e.g. Bode et al. 2000a; Jenkins et al. 2001). 
This will lead to an overestimate
of the total lensing probability for multiple images, 
$P(\Delta\theta)$, although the 
differential lensing probability normalized by $P(\Delta\theta)$, 
$[1/P(\Delta\theta)][dP(\Delta\theta)/d\Delta\theta]$, 
is probably less affected. In order to demonstrate the uncertainty,
we have used the modified PS mass function by Sheth \& Tormen (1999)
and the best fit parameters by Jenkins et al. (2001). 
It is found that the total lensing probabilities for small separation events 
in both CDM and WDM models are slightly reduced  as compared with
those predicted by the standard PS formalism. For example, 
at $z_{\rm s}=2$ and $\Delta\theta=0.1^{\prime\prime}$, the values of 
$P(\Delta\theta)$ in terms of the Sheth \& Tormen (1999) mass function 
are $1.2$ times smaller than those derived from
the PS approach, and this ratio is insensitive to cosmological models. 
The distribution of the differential lensing probability 
normalized by $P(\Delta\theta)$ based on the Sheth \& Tormen (1999) 
mass function remains roughly the same as in Figure 1.

Thirdly, the theoretically predicted distribution of image separations 
depends critically on the choice of the
cosmological parameters (e.g. Li \& Ostriker 2000). Therefore, 
a robust constraint on the WDM particle mass from
the study of lensing probability for multiple images with 
small angular separations needs {\it a priori} precise determinations 
of many other parameters such as $\Omega_{\rm M}$,
$\Omega_{\Lambda}$, $\sigma_8$ and $H_0$.

Finally, one may worry about 
the contamination of the supermassive black holes as lenses 
in the actual application of the gravitational
lensing probability to the observational tests of WDM model. Indeed, it is
believed that all the AGNs contain supermassive black holes with masses 
$M\sim10^8$ -- $10^9$ $M_{\odot}$ in their centers. 
The lensing effect by these 
supermassive black holes leads to an image separation of 
typically $\sim0.^{\prime\prime}1$ for a quasar at cosmological distance,
which is indeed of the same order of magnitude as
the effect produced by small galaxies discussed in the present {\sl Letter}.
Yet, the lensing probability of a background quasar due to the supermassive 
black holes residing in AGNs is considerably low ($\sim 10^{-8}$) 
(Augusto \& Wilkinson 2001).

Overall, we feel that gravitational lensing should serve as a critical
test for WDM model although further work will be needed to address in 
more detail various uncertainties in the issue.

\acknowledgments
We gratefully acknowledge the useful comments by an anonymous referee. 
This work was supported by
the National Science Foundation of China, under Grant No. 19725311
and the Ministry of Science and Technology of China, under Grant
No. NKBRSF G19990754.



\figcaption{The differential probabilities of image separations normalized
by the total  probabilities $P(\Delta\theta)$ for distant quasars 
at $z_{\rm s}=1$, 2 and 3 (from bottom to top) in CDM and
WDM models.   
\label{fig1}}

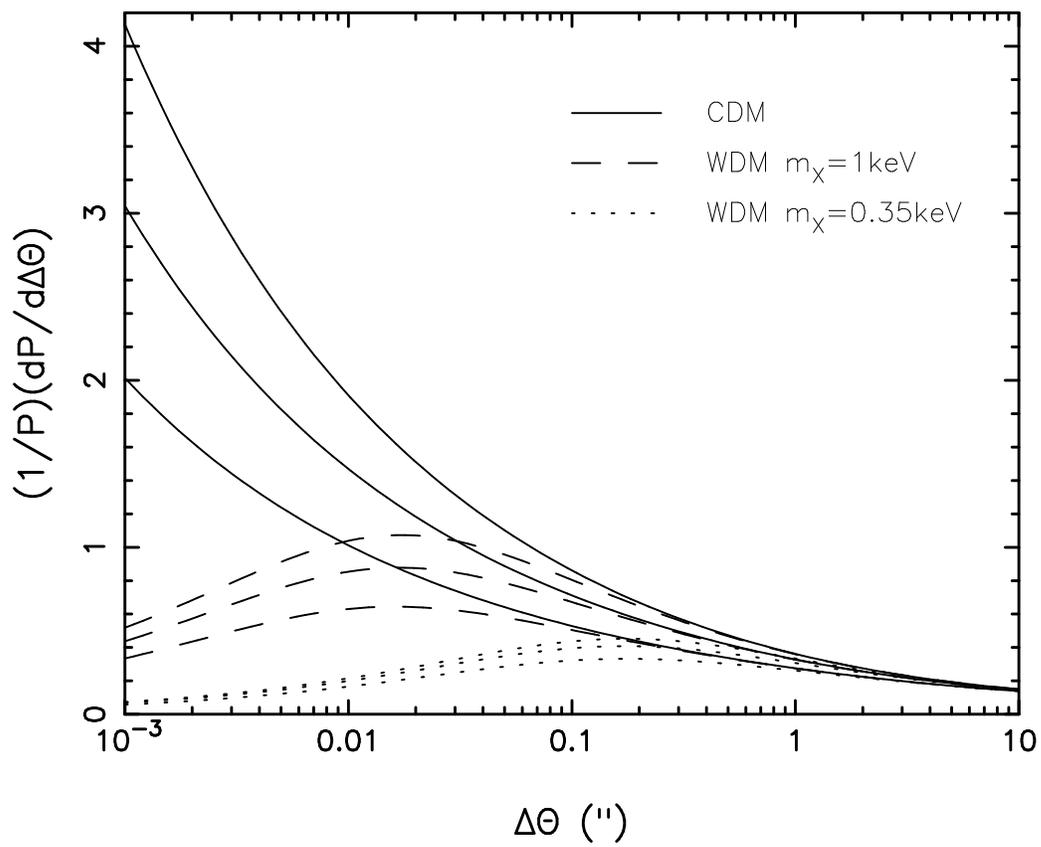
\end{document}